\documentclass[sigconf]{acmart}

\usepackage{enumitem}

\AtBeginDocument{%
  }

\setcopyright{acmlicensed}
\copyrightyear{2026}
\acmYear{2026}
\setcopyright{cc}
\setcctype{by}
\acmConference[RecSys '26]{20th ACM Conference on Recommender Systems}{September 27-October 02, 2026}{Minneapolis, MN, USA}
\acmBooktitle{20th ACM Conference on Recommender Systems (RecSys '26), September 27-October 02, 2026, Minneapolis, MN, USA}
\acmDOI{10.1145/3773078.3831759}
\acmISBN{979-8-4007-2284-4/2026/09}

\begin{document}

\title{Hierarchical GTV Estimation: Bridging the Gap between Container-Level Ranking and Item-Level Conversion}

\author{Zerong Lan}
\authornote{These authors contributed equally to this work.}
\email{lanzerong@meituan.com}
\orcid{0009-0006-8073-8399}
\affiliation{%
  \institution{Meituan}
  \city{Shanghai}
  \country{China}
}

\author{Fan Zhang}
\authornotemark[1]
\email{zhangfan133@meituan.com}
\orcid{0000-0002-9501-8478}
\affiliation{%
  \institution{Meituan}
  \city{Shanghai}
  \country{China}
}

\author{Chuang Chen}
\email{chenchuang08@meituan.com}
\orcid{0009-0004-7114-9213}
\affiliation{%
  \institution{Meituan}
  \city{Shanghai}
  \country{China}
}

\author{Tianchun Huang}
\email{huangtianchun@meituan.com}
\orcid{0009-0009-9422-0840}
\affiliation{%
  \institution{Meituan}
  \city{Shanghai}
  \country{China}
}

\author{Teng Zhang}
\authornote{Corresponding author.}
\email{zhangteng09@meituan.com}
\orcid{0009-0009-4199-2935}
\affiliation{%
  \institution{Meituan}
  \city{Shanghai}
  \country{China}
}

\author{Xingxing Wang}
\email{wangxingxing04@meituan.com}
\orcid{0000-0001-5495-0827}
\affiliation{%
  \institution{Meituan}
  \city{Shanghai}
  \country{China}
}

\renewcommand{\shortauthors}{Zerong et al.}

\begin{abstract}
Emerging e-commerce platforms, such as live-streaming and Online-to-Offline (O2O) services, inherently operate under a "Container-based" display paradigm. In this setting, a critical granularity mismatch exists: the platform performs ranking and estimation at the container level (e.g., livestream rooms or stores), whereas the actual transactions occur at the item level within these containers. This structural discrepancy complicates Gross Transaction Value (GTV) estimation, introducing unique challenges including high label variance, severe label sparsity, and dynamic item heterogeneity. To address these challenges, we propose a novel framework named \textbf{Hierarchical GTV Estimation} (HGE). HGE employs a \textbf{Hybrid Set-Aware Encoder} (HSAE) to model the dynamic composition of items and feature interactions. To mitigate label variance, we introduce a \textbf{Proxy-Label Learning} (PLL) strategy that decomposes post-click GTV into purchase probability, quantity, and unit price. Furthermore, a \textbf{Residual Correction and Ensemble Fusion} (RCEF) module is designed to handle label sparsity by effectively combining item-level aggregation with container-level inference. Extensive experiments on large-scale industrial datasets demonstrate that HGE significantly outperforms state-of-the-art baselines, achieving a $+0.63\%$ lift in XAUC offline and a $+1.42\%$ increase in Revenue Per Search (RPS) in online A/B tests. HGE framework has been fully deployed into our main traffic.
\end{abstract}


\keywords{Sponsored Search Advertising, GTV Estimation, Hierarchical Predicting}


\maketitle

\vspace{-0.5em}

\begin{figure}[t]
  \centering
  \includegraphics[width=0.95\linewidth]{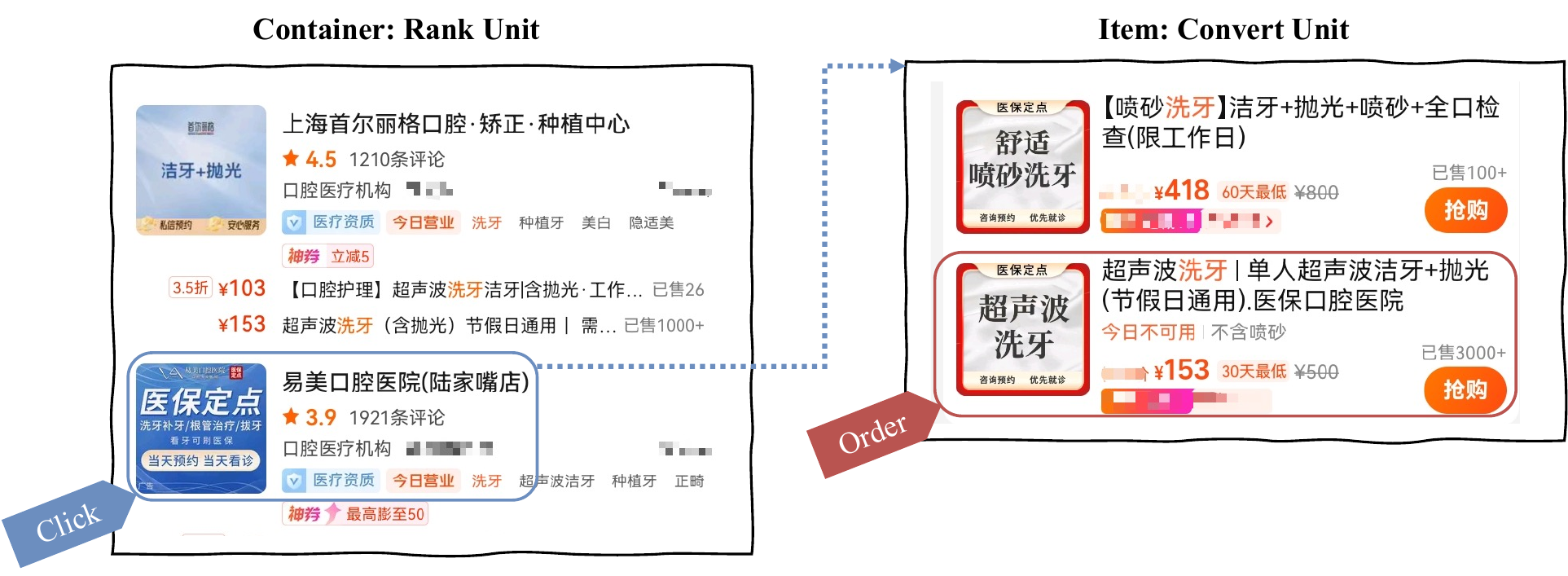}
  \vspace{-5pt} 
  \caption{Illustration of the hierarchical structure in sponsored search advertising.}
    \Description{The left side shows the Container level which is the target for ranking and bidding. The right side shows the Item level where the actual transactions occur. The core challenge is that the platform ranks containers, but the GTV is generated by a dynamic set of items with varying prices. A single observed transaction often fails to reflect the true expected value of the Container, leading to high variance and estimation bias.
    }
  \label{fig:container_item} 
  \vspace{-5pt} 
\end{figure}

\section{Introduction}

Sponsored search advertising serves as a fundamental revenue engine for modern e-commerce platforms, contributing a significant portion of the Gross Transaction Value (GTV)\cite{fan2019mobius}. Unlike traditional e-commerce where users directly search and click on specific products (Items)\cite{yu2025transun}, emerging platforms like live-streaming e-commerce (e.g., TikTok) and Online-to-Offline (O2O) services (e.g., Meituan) have shifted towards a "Container-based" display paradigm. In this setting, the user's decision process is inherently two-stage, as shown in Figure \ref{fig:container_item}. First, selecting a container (e.g., entering a livestream room or clicking a store card) and second, purchasing specific items (e.g., goods or vouchers) within that container. Consequently, the core objective of the advertising platform evolves from predicting the value of a single item to estimating the expected total GTV of a container, which is the aggregation of potential transactions across a dynamic set of items\cite{li2026delayed}.

To estimate the value of an advertisement, conventional methodologies typically employ point-wise prediction models, such as click-through rate (CTR) and conversion rate (CVR) estimators, combined with a deterministic pricing mechanism \cite{edelman2007internet,qiu2025unirom}. However, applying this formula directly to the container scenario is non-trivial. A naive solution is to treat the container as a "super-item" with an average price or to predict the conversion probability of the most popular item. While simple, these methods overlook the internal structure of the container. They fail to model the user's complex interaction with the diverse items inside, leading to suboptimal ranking decisions where high-potential containers are undervalued because their aggregate value is not fully captured. For instance, a model might assign a high prediction score to a container based on a high click probability, prompting the merchant to pay a premium for the traffic. However, if the user eventually purchases only a low-priced item (e.g., a cheap teeth cleaning service) rather than the expected high-value one (e.g., dental implants), the realized GTV will significantly lag behind the advertising cost, directly damaging the merchant's Return on Investment (ROI).

Developing a robust GTV estimation model for containers faces several unique challenges arising from the granularity mismatch between ranking (Container-level) and transaction (Item-level):

\textbf{High Variance and Label Noise}: The observed transaction value is highly stochastic. A user might purchase a low-priced item in a premium store simply due to momentary preference, even if the room has high-value potential. Using such a single-impression GTV as the ground truth label introduces extreme variance, confusing the model during training.
\textbf{Label Sparsity}: Transaction data at the item level are highly sparse. Conversion signals at the container level are already much sparser than click signals, and conversions for specific items within a store are extremely sparse, making direct modeling at the item level highly challenging. 
\textbf{Dynamics and Heterogeneity}: Containers are essentially dynamic sets of heterogeneous items. For instance, a store offers a mix of vouchers with varying discounts. Effectively modeling the interaction between Container-level features and Item-level features to predict the set-level value remains an open problem.

To tackle these challenges, we propose a novel framework named Hierarchical GTV Estimation (HGE), which bridges the gap between container-level predicting and item-level predicting. HGE models the GTV estimation process as a hierarchical probabilistic task. Specifically, we design a Hybrid Set-Aware Encoder (HSAE) to capture the dynamic composition of items within a container, allowing for fine-grained feature interaction between the container context and item attributes. To mitigate label variance, we introduce a Proxy-Label Learning (PLL) strategy that decomposes the post-click GTV into purchase probability, purchase quantity, and expected unit price, reducing the impact of outlier transactions. Furthermore, to address the severe label sparsity at the item level where conversion signals are extremely sparse. we propose a Residual Correction and Ensemble Fusion (RCEF) module. For samples with observed user behavior, RCEF aggregates item-level predictions to capture fine grained value signals. For samples lacking observed item interactions, the universal regression model is used to infer the potential container-level value. This hybrid approach ensures robust estimation throughout the entire traffic distribution, effectively overcoming the granularity mismatch between container-level supervision and item-level modeling.The main contributions of this work are summarized as follows:

1) We formally define the Container-Item hierarchical GTV estimation problem in O2O advertising, systematically identifying and analyzing three core challenges: high label variance in container-level GTV, severe label sparsity at the item-level conversion, and the granularity mismatch between container-level supervision and item-level feature modeling.

2) We propose HGE, a unified framework that addresses these challenges through three dedicated components: HSAE, PLL and RCEF.

3) Extensive experiments on large-scale industrial datasets from Meituan demonstrate that HGE significantly outperforms state-of-the-art baselines. We have obtained a +0.63pp lift in XAUC and +19.37pp lift in PCOC in offline experiment. Online A/B tests further confirm its effectiveness, achieving a +1.42\% lift in RPS.

  \vspace{-10pt} 

\section{Problem Formulation}
We focus on the post-click value estimation task in sponsored search. Let $\mathcal{U}$ and $\mathcal{P}$ be the sets of users and POIs (containers). Each POI $p \in \mathcal{P}$ contains a dynamic item set $\mathcal{I}_p = \{i_1, \dots, i_{K_p}\}$. The input feature $x$ comprises user $x_u$, POI $x_p$, context $x_c$, query $x_q$, and the top-K displayed items $\mathcal{I}_p$. A sample is a tuple $(x,y)$, where $y \in \mathbb{R}_{\geq0}$ is the realized GTV.
The goal is to estimate the expected GTV $\hat{y} = \mathbb{E}[y|x]$. The realized GTV $y$ depends on two sequential events: Conversion ($cvt \in \{0,1\}$) and Value Generation ($v \in \mathbb{R}_{\geq 0}$). Thus, the target is formulated as:
  \vspace{-10pt} 

\begin{equation}
\hat{y} = \mathbb{E}[y|x] = P(cvt=1|x)\mathbb{E}[v|cvt=1,x]
\end{equation}
where $\mathbb{E}[v|\cdot]$ implicitly models the quantity and price distribution. Our goal is to minimize the error between $y$ and $\hat{y}$ while handling the high variance of $v$.

\section{Methodology}
\subsection{Overall Architecture}
\begin{figure*}[t]
  \centering
  \includegraphics[width=0.85\linewidth]{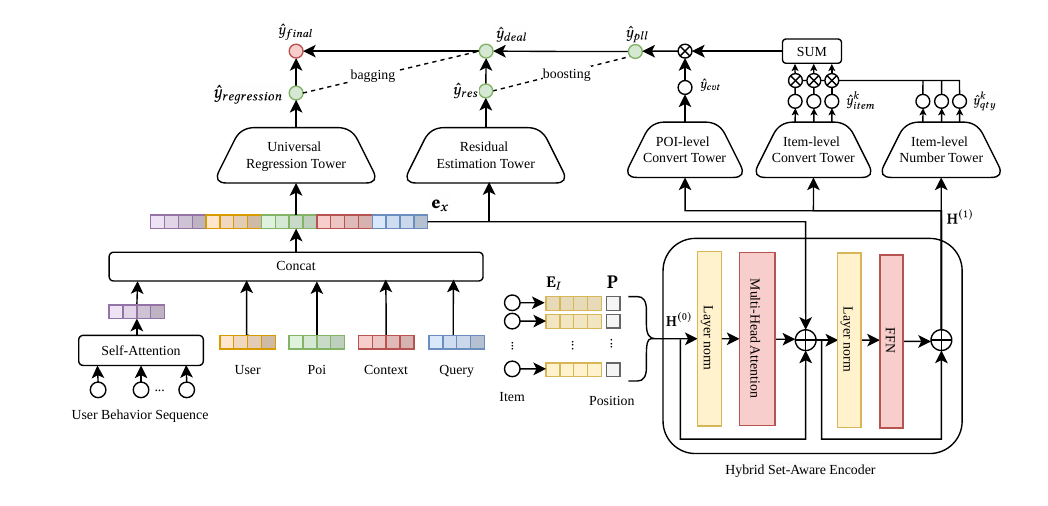}
  \vspace{-20pt} 
  \caption{Overall model architecture of HGE}
    \Description{}
  \label{fig:model_arch} 
  \vspace{-4pt} 
\end{figure*}

In this section, we present the proposed Hierarchical GTV Estimation (HGE) framework in detail. As illustrated in figure ~\ref{fig:model_arch}, HGE is designed to estimate the post-click expected GTV by effectively modeling the hierarchical structure of POIs and items. The framework consists of three core components: (1) Hybrid Set-Aware Encoder (HSAE) for capturing the interaction between the container context and the dynamic item set; (2) Proxy-Label Learning (PLL) for decomposing the high-variance GTV target into stable sub-tasks; and (3) Residual Correction and Ensemble Fusion (RCEF) for robust estimation by combining item-level aggregation and poi-level prediction.

The input feature space $x$ includes user features $x_u$, POI features $x_p$, context features $x_c$, query features $x_q$, and the sequence of top-K items $\mathcal{I}_p$. We first map these discrete features into dense embedding vectors. Let $\textbf{e}_u, \textbf{e}_p, \textbf{e}_c, \textbf{e}_q$ denote the embeddings for user, POI, context, and query, respectively. For the item set, we obtain a sequence of item embeddings $E_I \in \mathbb{R}^{K\times d_i}$ where $d_i$ is the embedding dimension of item.

\subsection{Hybrid Set-Aware Encoder}
A core challenge in container-level estimation is that the "value" of a POI is determined by its internal items, which form a dynamic and heterogeneous set. Traditional pooling methods (e.g., sum or average pooling) fail to capture the complex dependencies between the user's intent and specific items. To address this, we propose the \textbf{Hybrid Set-Aware Encoder (HSAE)} based on the Transformer architecture\cite{vaswani2017attention}.

\paragraph{Input Representation with Position Encoding.}
We first select the top‑k most frequently displayed items under each store as the candidate set for item‑level prediction, and initialize them into an item embedding matrix 
$\mathbf{E}_I \in \mathbb{R}^{K \times d}$. To enable the model to perceive the ranking order and relevance of these $K$ items, we incorporate position encodings $\mathbf{P} \in \mathbb{R}^{K \times d}$. The input representation for the HSAE module, denoted as $\mathbf{H}^{(0)}$, is obtained by element-wise addition:
\begin{equation}
    \mathbf{H}^{(0)} = \mathbf{E}_I + \mathbf{P}
\end{equation}
where $\mathbf{P} = [\mathbf{p}_1; \mathbf{p}_2; \dots; \mathbf{p}_K]$ represents the position embeddings for the 1-st to $K$-th positions.

\paragraph{Hybrid Set-wise Transformer}
To capture the complex interactions among items, we feed $\mathbf{H}^{(0)}$ into a Transformer encoder layer. 
To enhance the deeper competitiveness modeling, we concatenate the output of the Multi-Head Self-Attention (MHSA) layer with poi features, user features, and current contextual information, and then feed the enriched representation into the subsequent Feed‑Forward Network (FFN). The timing of introducing this information presents a trade-off: early introduction dilutes competitive signals, while complete omission prevents modeling competition differences across user-poi-query triplets. Our solution is to inject this information after the MHSA layer. This allows the model to first learn robust base representations, then specifically modulate them with competitive context, achieving both objectives.

Residual connections\cite{he2016deep} and Layer Normalization (LN)\cite{ba2016layer} are applied after each sub-layer. 
The calculation process is shown in Eq.3-4, where $\textbf{e}_x = concat([\textbf{e}_u, \textbf{e}_p, \textbf{e}_c, \textbf{e}_q])$.
\begin{align}
    \mathbf{H}' &= \text{LN}\left(concat(\mathbf{H}^{(0)} + \text{MHSA}(\mathbf{H}^{(0)}), \mathbf{e}_{x})\right)
\end{align}
\vspace{-12pt}
\begin{align}
    \mathbf{H}^{(1)} &= \text{LN}\left(\mathbf{H}' + \text{FFN}(\mathbf{H}')\right)
\end{align}
 \vspace{-20pt}

\subsection{Proxy-Label Learning (PLL)}
Directly regressing the total GTV $y$ is suboptimal due to the severe label sparsity and high variance issues discussed in Section 1. To mitigate this, we propose the \textbf{Proxy-Label Learning (PLL)} strategy, which decomposes the GTV estimation task into three hierarchical sub-tasks: (1) POI-level conversion probability, (2) Item-level purchase probability, and (3) Item-level purchase quantity.
      
\paragraph{GTV Decomposition}
Based on the sequential decision process of users, we formulate the expected GTV $\hat{y}$ as the product of the probability of any conversion occurring in the POI and the expected value generated from specific items given that conversion. The decomposition formula is defined as:
\begin{equation}
    \hat{y}_{pll} = \hat{y}_{cvt} \times \sum_{k=1}^K \left( \hat{y}_{item}^{k} \times \hat{y}_{qty}^{k} \times \text{Price}_k \right)
\end{equation}
where:
\begin{itemize}[leftmargin=*, noitemsep, topsep=0pt, partopsep=0pt, parsep=0pt]
    \item $\hat{y}_{cvt} \in [0, 1]$ is the probability that the user makes \textit{any} purchase within the top-$K$ items of the POI.
    \item $\hat{y}_{item}^{k} \in [0, 1]$ is the conditional probability that the user purchases the $k$-th item, given that a conversion event occurs in the POI.
    \item $\hat{y}_{qty}^{k} \in \mathbb{R}^{+}$ is the expected quantity of the $k$-th item purchased.
    \item $\text{Price}_k$ is the deterministic unit price of the $k$-th item.
\end{itemize}

\paragraph{Prediction Towers}
To estimate these components, we design three specific prediction heads. The \textbf{POI-level Convert Tower} estimates the $\hat{y}_{cvt}$. We employ a Multi-Layer Perceptron (MLP) taking the mean pooling representation $\mathbf{z}_{poi} = mean(\mathbf{H}^{(1)})$ (from Eq. 4) as input:
\begin{equation}
    \hat{y}_{cvt} = \sigma(\text{MLP}_{global}(concat([\mathbf{z}_{poi}, \textbf{e}_x])))
\end{equation}
where $\sigma(\cdot)$ denotes the sigmoid activation function. 
The \textbf{Item-level Convert Tower} and the \textbf{Item-level Number Tower} estimates the $\hat{y}_{item}^{k}$ and $\hat{y}_{qty}^{k}$ respectively. We leverage the context-aware item embeddings $\mathbf{H}^{(1)} = [\mathbf{h}_1, \dots, \mathbf{h}_K]$ generated by the HSAE module. Crucially, we employ \textit{shared} MLPs to process each item embedding, ensuring parameter efficiency and generalization across different items. For the $k$-th item:
\begin{align}
    \hat{y}_{item}^{k} &= \sigma(\text{MLP}_{item\_cvr}(concat([\mathbf{h}_k, \textbf{e}_x])))
\end{align}
\vspace{-10pt}
\begin{align}
    \hat{y}_{qty}^{k} &= \text{Softplus}(\text{MLP}_{item\_qty}(concat([\mathbf{h}_k, \textbf{e}_x])))
\end{align}
Here, we use the $\text{Softplus}(x) = \ln(1+e^x)$ activation function for the quantity head to strictly constrain the predicted number to be positive.

\paragraph{Variance Reduction Justification}
The benefit of decomposition can be formally understood via the Law of Total Variance. Let $N$ denote the total purchase quantity (so $y = N \cdot \text{Price}$). Since $N$ is realized only upon conversion, we have:
\begin{equation}
    \text{Var}(N) = \mathbb{E}[\text{Var}(N \mid cvt)] + \text{Var}(\mathbb{E}[N \mid cvt])
\end{equation}
Because $cvt$ is Bernoulli and the conditional purchase quantity concentrates at low values ($\leq 2$ in over 90\% of conversions), both terms on the right-hand side are substantially smaller than directly regressing $N$ as a single zero-inflated count target. By decomposing GTV into $\hat{y}_{cvt}$, $\hat{y}_{item}^k$, and $\hat{y}_{qty}^k$, PLL replaces one high-variance regression with multiple low-variance sub-tasks, each constrained to a well-behaved output range ($[0,1]$ for probabilities, $\mathbb{R}^+$ for quantity).

By explicitly modeling the purchase probability and quantity for each item, PLL also allows the model to differentiate between high-price, low-probability items and low-price, high-probability items, leading to a more robust total GTV estimation.

\subsection{Residual Correction and Ensemble Fusion}
\textbf{Residual Correction and Ensemble Fusion (RCEF)}  consists of residual learning and ensemble fusion.
First, considering that not all in‑store orders originate from the top‑k items, we introduce a lightweight residual tower $\mathcal{T}_{res}$ to correct local biases from the PLL module. 
\begin{align}
    \hat{y}_{res} = \mathcal{T}_{res}(\mathbf{e}_{x}). 
\end{align}
Following the residual correction paradigm, the refined prediction is $\hat{y}_{deal} = \hat{y}_{pll} + \hat{y}_{res}$, where $\hat{y}_{pll}$ is the prediction from the PLL module. 

Second, due to the extreme sparsity of conversion behavior at the item level, we enhance the stability of the estimation by combining $\hat{y}_{deal}$ with the output $\hat{y}_{regression}$ from the \textbf{Universal Regression Tower} through a weighted summation: 
\begin{align}
    \hat{y}_{final} = \alpha \cdot \hat{y}_{deal} + (1 - \alpha) \cdot \hat{y}_{regression}, \alpha \in [0, 1]
\end{align}
The \textbf{Universal Regression Tower} can be instantiated with any generic regression architecture. In this work, we implement it using a combination of DIN\cite{zhou2018deep} and DCN\cite{wang2021dcn}.

\paragraph{Learning Objective}
We adopt a multi-task loss: 
\begin{align}
\mathcal{L}_{total} = \mathcal{L}_{regression} + \lambda_1 \mathcal{L}_{cvt} + \lambda_2 \mathcal{L}_{item} + \lambda_3 \mathcal{L}_{qty}
\end{align}
The Huber loss (Eq. 13) is employed for the regression tasks, i.e., the universal regression prediction tower $\mathcal{L}_{reg}$ and the item-level quantity prediction tower $\mathcal{L}_{qty}$, with $\delta$ as a hyperparameter threshold. The binary cross-entropy (BCE) loss is used for two binary classification tasks: $\mathcal{L}_{cvt}$ for predicting order occurrence at the POI's top-k item set level, and $\mathcal{L}_{item}$ for predicting the conversion of the k-th specific item, respectively.
\begin{equation}
    L_{\delta}(y, \hat{y}) = 
        \begin{cases} 
            \frac{1}{2}(y - \hat{y})^2 & \text{if } |y - \hat{y}| \le \delta \\
            \delta \cdot |y - \hat{y}| - \frac{1}{2}\delta^2 & \text{otherwise}
        \end{cases}
\end{equation}

\section{Experiments}

\subsection{Experiment Settings}

\paragraph{Datasets}
The offline experiments were conducted using large-scale industrial datasets collected from the Meituan advertising system log. The total dataset contains 712M records, including 121M users and 949K POIs. To prevent look-ahead bias, we adopt a strict Out-of-Time (OOT) chronological split: the last 5\% of records by time are reserved for testing (29M records, 13M users, 408K POIs), while the remaining 95\% (683M records) are used for training.

\paragraph{Evaluation Metrics}
To comprehensively evaluate the model performance, we adopt \textbf{XAUC} as the primary metric, measuring how the ranking induced by the predicted GTV is in agreement with the ideal ranking\cite{zhan2022deconfounding}. 
\begin{equation}
    \text{XAUC} = \frac{1}{|\mathcal{P}|} \sum_{(i, j) \in \mathcal{P}} \mathbb{I}\left( \text{sign}(y_i - y_j) = \text{sign}(\hat{y}_i - \hat{y}_j) \right)
\end{equation}
where $\mathcal{P}$ denotes the sample pairs, $\mathbb{I}(\cdot)$ denotes the indicator function, and $\text{sign}(\cdot)$ denotes sign function.

Additionally, we use \textbf{PCOC} (Predicted Click-Over-Click)\cite{niculescu2005predicting} to measure the calibration ratio between the predicted and actual GTV. 
To further assess the estimation accuracy at the merchant level, we introduce the typical measurement for evaluating continuous-value-regression accuracy \textbf{MAE}\cite{lin2023tree}.

\paragraph{Baselines}
We compare HGE against widely recognized baselines: 
\textbf{WLR}, This method transforms GTV regression into a weighted binary classification task, where the predicted odds from the logistic model approximate the expected GTV \cite{covington2016deep}. \textbf{DLRM}, a standard deep learning regression\cite{naumov2019deep} method with DIN \cite{zhou2018deep} for sequence modeling and DCN\cite{wang2021dcn} for feature crossing. \textbf{RankMixer}, which replaces self-attention with token mixing for efficiency\cite{zhu2025rankmixer}; and \textbf{CREAD}, a calibration-focused framework transforming continuous estimation into multi-classification task \cite{sun2024cread}.

\paragraph{Training Details}
We set the top-$K$ item number to $K=10$. The embedding dimension is $8$ for all categorical features. The HSAE module uses a single Transformer encoder layer with hidden dimension $d=256$. The MLP towers use hidden layers of $[1024, 512, 256]$ with Dice activation. The Universal Regression Tower combines DIN and DCN with hidden layers $[128, 32]$. The model is trained with Adam optimizer (learning rate $9\times10^{-4}$, batch size $6{,}000$) for one epoch over the training set. Negative sampling with a rate of $0.3$ is applied to balance the positive-negative ratio. For the multi-task loss (Eq.~12), we set $\lambda_1=0.3$, $\lambda_2=0.03$, $\lambda_3=0.03$, and the Huber loss threshold $\delta=1.0$. The ensemble weight $\alpha$ in RCEF is implemented as a per-sample learnable parameter via a lightweight gating network conditioned on user-POI-context features, rather than a fixed scalar.

\subsection{Experiment Results}

\paragraph{Overall Performance}
The performance comparison of different methods is reported in Table~\ref{tab:main_results}. 
From the results, we observe that our proposed HGE framework achieves state-of-the-art performance across most metrics. 
Specifically, HGE yields a significant improvement in \textbf{XAUC} (0.8158), outperforming the strongest continuous value prediction baseline CREAD and the token-based baseline RankMixer. This demonstrates HGE's superiority in ranking value-sensitive instances. 

Regarding calibration, HGE achieves a \textbf{PCOC} of 0.9858, which is remarkably close to the ideal value of 1.0, indicating that the predicted GTV is highly aligned with the actual value. 
These improvements validate that our hierarchical modeling strategy (PLL) and residual correction (RCEF) effectively mitigate the estimation bias and variance inherent in GTV estimation.

\begin{table}[t]
  \caption{Performance comparison on the offline dataset. The best results are highlighted in bold.}
  \label{tab:main_results}
  \centering
  \footnotesize 
  \renewcommand{\arraystretch}{0.9} 
  \setlength{\tabcolsep}{4mm}
  \begin{tabular}{l|ccc}
    \toprule
    \textbf{Method} & \textbf{XAUC} $\uparrow$ & \textbf{PCOC} & \textbf{MAE} $\downarrow$ \\
    \midrule
    WLR       & 0.8043 & 0.7663 & 5.1453 \\
    DLRM      & 0.8051 & 0.7787 & 5.0453 \\
    RankMixer & 0.8105 & 0.7528 & 5.1735 \\
    CREAD     & 0.8095 & 0.7921 & 5.0818 \\
    \midrule
    \textbf{HGE (Ours)} & \textbf{0.8158} & \textbf{0.9858} & \textbf{4.9915}\\
    \bottomrule
  \end{tabular}
\end{table}

\paragraph{Ablation Study}
Table~\ref{tab:ablation} shows the contribution of each component. Removing \textbf{PLL} causes the most drastic drop (XAUC 0.8158 $\to$ 0.7873), confirming that decomposing high-variance targets is crucial. Removing \textbf{HSAE} also degrades performance, highlighting the need for set-level modeling over simple pooling. Finally, removing \textbf{RCEF} causes a slight decline in XAUC. Compared to the complete model, the higher MAE of w/o RCEF group indicates that the GTV results obtained by the RCEF module are closer to the true label values, proving the residual tower's effectiveness in bias correction.

\paragraph{Online A/B Test}
We deployed HGE to the online production environment and conducted a 13-day A/B test, following a prior A/A validation phase to ensure baseline equivalence. HGE achieved a $+1.42\%$ improvement in Revenue Per Search (RPS) with a 95\% confidence interval of $[+0.80\%, +2.05\%]$, which excludes zero and confirms statistical significance. This increase is highly meaningful on mainstream online traffic and supports full-scale traffic expansion.

\begin{table}[h]
  \caption{Ablation study of different components in HGE.}
  \label{tab:ablation}
  \centering
  \footnotesize 
  \renewcommand{\arraystretch}{0.9} 
    \vspace{-10pt} 
  \setlength{\tabcolsep}{3mm}
  \begin{tabular}{l|ccc}
    \toprule
    \textbf{Variant} & \textbf{XAUC} $\uparrow$ & \textbf{PCOC} & \textbf{MAE} $\downarrow$ \\
    \midrule
    \textbf{Full Model} & \textbf{0.8158} & 0.9858 & \textbf{4.9915} \\
    w/o HSAE      & 0.8031 & 1.1012 & 5.25862 \\
    w/o PLL      & 0.7873 & 0.4180 & 6.98752 \\
    w/o RCEF     & 0.8146 & \textbf{0.9960} & 5.01299 \\
    \bottomrule
  \end{tabular}
  \vspace{-10pt} 
\end{table}

\section{Conclusion}

This paper addresses the granularity mismatch problem in container-based GTV estimation by proposing a Hierarchical GTV Estimation (HGE) framework. HGE incorporates a Hybrid Set-Aware Encoder (HSAE) to model the semantic information of dynamic item sets within containers. A Proxy-Label Learning (PLL) strategy is introduced to decompose the high-variance regression target into stable subtasks, enabling fine-grained user preference learning. Furthermore, a Residual Correction and Ensemble Fusion (RCEF) module refines predictions by combining the stability of a universal regression model with the precision of a residual tower. Experiments on industrial offline and online datasets show that HGE significantly outperforms state-of-the-art methods in both ranking accuracy (XAUC) and calibration capability (PCOC). Successful fully deployment in Meituan's advertising system confirms the framework's practical value in improving platform revenue.

\bibliographystyle{unsrt}
\bibliography{recsys_2026_short}

\end{document}